\documentstyle[aps,cite,enumerate]{revtex}

\begin{document}
\draft
\title{Noncyclic geometric phase for neutrino oscillation}
\author{Xiang-Bin Wang
\footnote{E-mail address: scip7236@leonis.nus.edu.sg }, L.C. Kwek, Yong Liu
\footnote{Email address: phyliuy@nus.edu.sg} and C.H. Oh
\footnote{E-mail address: phyohch@leonis.nus.edu.sg}}
\address{Department of Physics,Faculty of Science, National University
of Singapore, Lower Kent Ridge Road, Singapore 119260, Republic of
Singapore.}

\maketitle
\newcommand{\rivpt}{\rule{.4pt}{4pt}}
\newcommand{\leftmk}{\makebox[8cm]{\hrulefill\rivpt}}
\newcommand{\rightmk}{\hspace{9cm}\rivpt\raisebox{1ex}{\makebox[8cm]{\hrulefill}}}

\begin{abstract}  We provide explicit formulae for
the noncyclic geometric phases or Pancharatnam phases of neutrino
oscillations. Since Pancharatnam phase is a generalization of the
Berry phase, our results generalize the previous findings for
Berry phase in a recent paper [Phys. Lett. B, 466 (1999) 262].
Unlike the Berry phase, the noncyclic geometric phase offers
distinctive advantage in terms of measurement and prediction. In
particular, for three-flavor mixing, our explicit formula offers
an alternative means of determining the CP-violating phase. Our
results can also be extended easily to explore geometric phase
associated with neutron-antineutron oscillations.
\end{abstract}

\section{Introduction}

Pontecorvo's suggestion \cite{pontecorvo} nearly half a century
ago that neutrinos had finite masses implied that neutrino mass
eigenstates need not be identical with the weak eigenstates and
thus may give rise to neutrino oscillations.  Indeed, recent
experiments from atmospheric neutrino data in the Super-Kamiokande
experiments \cite{fukuda}, IMB collaboration \cite{imb}, Soudan II
\cite{soudan} and MACRO \cite {macro} experiments have provided
strong confirmation of such oscillations.

In a recent paper\cite{blasone}, it was found that the geometric
phase appears naturally in the standard Pontecorvo formulation of
neutrino oscillations. The Berry phase \cite{berry} for
oscillating neutrinos was calculated and found to be a functional
of the mixing angle for the two-flavor neutrinos. Since it is
possible in principle to observe the geometric phase, it was
suggested that the mixing angle could then be deduced through the
observation of the Berry phase. However, the measurement of the
Berry phase is only applicable for cyclic adiabatic evolution.
Thus one can only measure a state after it has undergone a closed
circuit with some period, $T$. For neutrinos, this period is
relatively long. Thus in order to measure the Berry phase for
neutrinos, we need to place the detector sufficiently distant from
the source so that the neutrinos traverse exact distance
corresponding to a complete cycle. Experimentally, this technique
can be difficult.

Three-flavor neutrino oscillations are also particularly
interesting due to its physical implications in CP violation.
Nevertheless, based on the formula in ref \cite{blasone}, it may
not be easy to determine the CP-violating phase. In this paper, we
generalize the idea of the Berry phase to a non-cyclic geometric
phase and discuss how some of the above difficulties could be
circumvented through the generalization. Naturally, our explicit
formula for the non-cyclic geometric phase reduces to the Berry
phase formula in ref \cite{blasone} when the time of measurement
is set to the oscillating period of the neutrino.

The generalization of geometric phase to noncyclic
evolution\cite{wagh1,wu} dates back to an important seminal paper
by Pancharatnam\cite{pancha}. Experimental results for non-cyclic
geometric phase or Pancharatnam phase have been demonstrated
recently in experiments \cite{wagh2,wein}. Following the idea
raised in ref. \cite{blasone}, one can in principle extract
information concerning states of the neutrinos by observing the
noncyclic geometric phase at different times.

This paper is organized as follows.  In section \ref{noncyclic},
we briefly describe the notion of non-cyclic phase and consider
two-flavor neutrino oscillation.  In section \ref{three}, we
extend the same calculation to the three-flavor case and show how
the CP-violating phase can in principle be deduced from the
non-cyclic phase. Finally, we summarizes the results in section
\ref{discuss}.

\section{Noncyclic geometric phase}\label{noncyclic}

We first explain how we can compute the non-cyclic geometric phase
\cite{mendas,polavieja}. Suppose state $|\chi(0)\rangle$ evolves
to a state $|\chi(t)\rangle$ after a certain time $t$. If the
scalar product $$\langle
\chi(0)|\exp\left[\frac{i}{\hbar}\int^t_0<E>(t')dt'\right]|\chi(t)\rangle$$
can be written as $r\exp[i\beta]$, where $r$ is a real number,
then we say that the non-cyclic geometric phase due to the
evolution from $|\chi(0)\rangle$ to $|\chi(t)\rangle$ is the
$\beta$. This non-cyclic geometric phase generalizes the cyclic
geometric phase since the latter can be regarded as a special case
of the former for which $r=1$.

We first consider the two-flavor oscillating neutrino states as
\begin{eqnarray}
|\nu_e(0)\rangle & = & \cos\theta
|\nu_1\rangle+\sin\theta|\nu_2\rangle \\ |\nu_{\mu}(0)\rangle & =
& -\sin\theta|\nu_1\rangle+\cos\theta|\nu_2\rangle
\end{eqnarray}
respectively. At time $t$, the states $|\nu_e\rangle$ and
$|\nu_{\mu}\rangle$ evolve to the states
\begin{eqnarray}
|\nu_e(t)\rangle & = & e^{-iHt}|\nu_e(0)\rangle=e^{-i\omega_1
t}\cos\theta|\nu_1\rangle +e^{-i\omega_2 t}\sin\theta|\nu_2\rangle
\\ |\nu_{\mu}(t)\rangle & = &
e^{-iHt}|\nu_{\mu}(0)\rangle=-e^{-i\omega_1
t}\sin\theta|\nu_1\rangle +e^{-i\omega_2 t}\cos\theta|\nu_2\rangle
\end{eqnarray}
To calculate the noncyclic geometric phase for the evolution from
$|\nu_e(0)\rangle$ to $|\nu_e(t)\rangle$, we define a new state,
$|\tilde{\nu}_e(t)\rangle$, given by
\begin{eqnarray}
|\tilde{\nu}_e(t)\rangle & = &
\exp\left[i\int^t_0<E>(t')dt'\right]|\nu_e(t)\rangle \\
 & = &
\exp[i(\omega_1\cos^2\theta+\omega_2\sin^2\theta)t]|\nu_e(t)\rangle
\end{eqnarray}
so that
\begin{eqnarray}
\langle\nu_e(0)|\tilde{\nu}_e(t)\rangle & = &
\exp[i(\omega_1\cos^2\theta+\omega_2\sin^2\theta)t]
\left[\cos^2\theta e^{-i\omega_1t}+\sin^2\theta
e^{-i\omega_2t}\right]\\ & \equiv & r\exp[i\beta]
\end{eqnarray}
Denoting $\displaystyle \Omega=-\frac{\omega_1+\omega_2}{2}$ and
$\displaystyle \phi=-\frac{\omega_1-\omega_2}{2}$, we have
\begin{equation}
\langle\nu_e(0)|\tilde{\nu}_e(t)\rangle=\exp[i(\omega_1\cos^2\theta+\omega_2\sin^2\theta+
\Omega) t] \left[\cos^2\theta e^{i\phi t}+\sin^2\theta e^{-i\phi
t}\right]
\end{equation}
This can also be written as
\begin{eqnarray}
\langle\nu_e(0)|\tilde\nu_e(t)\rangle=\exp[-i\phi t \cos2\theta]
\left[\cos^2\theta e^{i\phi t}+\sin^2 \theta e^{-i\phi t}\right]
\equiv re^{i\beta}
\end{eqnarray}
where the explicit expressions for $r$ and $\beta$ are then given
by
\begin{eqnarray}
r=\sqrt{1-\sin^22\theta\sin^2\phi t}
\end{eqnarray}
and
\begin{eqnarray}
\beta=- \phi t \cos2\theta +\tan^{-1}[\cos 2\theta\tan(\phi
t)].\label{phase1}
\end{eqnarray}
We can see that there is indeed a nonzero geometric phase under
non-cyclic evolution. In particular, this phase reduces to the
value of $2\pi \sin^2\theta$ if the time $t$ is set to the period
of the oscillating neutrinos, that is $\displaystyle
t=\frac{2\pi}{\omega_2-\omega_1}$, upon choosing the appropriate
branch. Thus one recovers the result in ref \cite{blasone} for the
Berry phase. However, since the noncyclic geometric phase can be
measured at arbitrary time $t$, there is no need to restrict the
time, $t$, of the measurement to exactly one period,
$T=\frac{2\pi}{\omega_2-\omega_1}$. From the experimental point of
view, such relaxation would facilitate the measurement of the
geometric phase. Hence, from eq(\ref{phase1}), one can see that it
is possible in principle to deduce the mixing angle either by
measuring the value of $r$ (which can be done by counting the
neutrino flux) or detecting the geometric phase at two different
times and then solving the resulting simultaneous equations for
$\theta$ and $\phi t$.

It is also possible to compute the other components, namely
$\langle\nu_e(0)|\tilde\nu_{\mu}(t)\rangle$,
$\langle\nu_{\mu}(0)|\tilde\nu_e(t)\rangle$ and
$\langle\nu_{\mu}(0)|\tilde\nu_{\mu}(t)\rangle$ and their
associated noncyclic geometric phases.  The results are summarized
in the following tables.

\begin{tabular}{l|l}
\hline  component & expression \\ \hline
$\langle\nu_e(0)|\tilde\nu_{\mu}(t)\rangle$ & $(\exp[-2 i \phi t
\sin^2 \theta ]-\exp[2 i\phi t \cos^2 \theta ]) \cos \theta \sin
\theta$ \\ $\langle\nu_{\mu}(0)|\tilde\nu_e(t)\rangle$  & $(\exp[-
2 i \phi t \cos^2 \theta ]-\exp[2 i\phi t \sin^2 \theta ]) \cos
\theta \sin \theta$ \\
$\langle\nu_{\mu}(0)|\tilde\nu_{\mu}(t)\rangle$ & $\exp[i\phi t
\cos2\theta] \left[\cos^2\theta e^{-i\phi t}+\sin^2e^{i\phi
t}\right]$\\ \hline

\end{tabular}

\begin{tabular}{l|l|l}
\hline component & $r$ & $\beta$ \\ \hline
$\langle\nu_e(0)|\tilde\nu_{\mu}(t)\rangle$ & $\sin 2 \theta \sin
\phi t$ & $ \displaystyle \phi t \cos 2 \theta - \frac{\pi}{2}$ \\
$\langle\nu_{\mu}(0)|\tilde\nu_e(t)\rangle$ & $\sin 2 \theta \sin
\phi t$ & $ \displaystyle -\phi t \cos 2 \theta - \frac{\pi}{2}$
\\
$\langle\nu_{\mu}(0)|\tilde\nu_{\mu}(t)\rangle$ & $
r=\sqrt{1-\sin^22\theta\sin^2\phi t}$ &  $\phi t \cos2\theta -
\tan^{-1}[\cos 2\theta\tan(\phi t)]$ \\ \hline
\end{tabular}

\section{Three-flavor Oscillation}\label{three}
The noncyclic geometric phase for the case of three-flavor mixing
can also be computed using the same method. In the case of
three-flavor mixing, the electron neutrino state  at time $t$ is
\begin{eqnarray}
|\nu_e(t)\rangle=e^{-i\omega_1 t}\cos\theta_{12}\cos\theta_{13}
|\nu_1\rangle+e^{-i\omega_2
t}\sin\theta_{12}\cos\theta_{13}|\nu_2\rangle \nonumber \\
 +e^{-i\omega_3
t}e^{i\delta}\sin\theta_{13}|\nu_3\rangle
\end{eqnarray}
where $\theta_{12}$ and $\theta_{13}$ are the appropriate mixing
angles. More generally, one can consider the mixing in terms of
the Cabibbo-Maskawa-Kobayashi (CKM) matrix,$U$, using the
parametrization
\begin{equation}
U = \left( \begin{array}{ccc} c_{12} c_{13} & s_{12} c_{13} & e^{i
\delta} s_{13}
\\
- s_{12} c_{23} - c_{12} s_{23} s_{13} e^{-i \delta} & c_{12}
c_{23} - s_{12} s_{23} s_{13} e^{-i \delta} & s_{23} c_{13}
\\
s_{12} s_{23} - c_{12} c_{23} s_{13} e^{-i \delta} & - c_{12}
s_{23} - s_{12} c_{23} s_{13} e^{-i \delta} & c_{23} c_{13}
\end{array}
\right)
\end{equation}
where $s_{ij}= \sin \theta_{ij}$, $c_{ij} = \cos \theta_{ij}$ and
$\delta$ is the CP violating phase of the CKM matrix.

Moreover, as in the two-flavor case, for the electron neutrino,
\begin{eqnarray}
|\tilde{\nu}_e(t)\rangle =
\exp[i\int^t_0<E>(t')dt']|\nu_e(t)\rangle
\end{eqnarray}
with
$$<E>(t)=\omega_1\cos^2\theta_{12}\cos^2\theta_{13}+\omega_2\sin^2\theta_{12}\cos^2\theta_{13}
+\omega_3\sin^2\theta_{13}.$$ A straightforward calculation yields
\begin{eqnarray}
\langle\nu_e(0)|\tilde \nu_e(t)\rangle & = & \exp[i
(\omega_1\cos^2\theta_{12}\cos^2\theta_{13}+\omega_2\sin^2\theta_{12}\cos^2\theta_{13}
+ e^{i \delta} \omega_3\sin^2\theta_{13})t] \nonumber \\ & &
\left[\cos^2\theta_{12}\cos^2\theta_{13}e^{-i\omega_1t}
+\sin^2\theta_{12}\cos^2\theta_{13}e^{-i\omega_2t}+  e^{2 i
\delta} \sin^2\theta_{13}e^{-i\omega_3t}\right] \nonumber \\
&\equiv & r_{ee} e^{i\beta_{ee}}\label{inner1}
\end{eqnarray}
where $r_{ee}$ and $\beta_{ee}$ are the modulus and phase of the
inner product in eq(\ref{inner1}) between $\nu_e-\nu_e$ states
respectively. The left hand side of eq(\ref{inner1}) can be
written as
\begin{eqnarray}\langle\nu_e(0)|\tilde \nu_e(t)\rangle& = &
\exp[i(\omega_1\cos^2\theta_{12}\cos^2\theta_{13}+
\omega_2\sin^2\theta_{12}\cos^2\theta_{13} +
e^{i\delta}\omega_3\sin^2\theta_{13}-\frac{\omega_1+\omega_2}{2})t]
\nonumber
\\ & & \times\left[\cos^2\theta_{12}\cos^2\theta_{13}e^{i\phi t}
+\sin^2\theta_{12}\cos^2\theta_{13}e^{-i\phi t}+  e^{2 i \delta}
\sin^2\theta_{13}e^{i(2q- 1)\phi t}\right] \nonumber \\ & = &
\exp[ \omega_1 \sin \theta_{13}^2 (e^{2 i \delta} -1) + i(2\phi
\sin^2\theta_{12}\cos^2\theta_{13}+2q\phi  e^{i \delta}
\sin^2\theta_{13}-\phi)t] \nonumber \\ & & \times \left[
\cos^2\theta_{12}\cos^2\theta_{13}e^{i\phi t}
+\sin^2\theta_{12}\cos^2\theta_{13}e^{-i\phi t}+\sin^2\theta_{13}
e^{2 i \delta} e^{i(2q -1)\phi t}\right]\end{eqnarray} where
$\displaystyle q=\frac{\omega_3-\omega_1}{\omega_2-\omega_1}$ and
$\displaystyle \phi= -\frac{\omega_1- \omega_2}{2}$ as defined
previously for the two-flavor case. After some algebraic
manipulations, the geometric phase can be found to be
\begin{eqnarray} \beta_{ee} \equiv \beta & = & \omega_1 \sin \theta_{13}^2 (\cos( 2 \delta) -1) +
(2 \phi \sin^2\theta_{12}\cos^2\theta_{13}+2q\phi \cos( 2 \delta)
\sin^2\theta_{13}-\phi)t \nonumber \\ & &  +\tan^{-1}\frac{\cos
2\theta_{12}\cos^2\theta_{13}\sin\phi t-\sin^2\theta_{13}\sin[
(2q-1)\phi t - 2 \delta]} {\cos^2\theta_{13}\cos\phi t
+\sin^2\theta_{13}\cos[(2q-1)\phi t - 2 \delta ] } \label{cp}
\end{eqnarray}
In general, we do not expect the CP-violating phase, $\delta$, to
be zero. However if we take the CP violating phase to be zero, as
in ref \cite{blasone}, then we get
\begin{eqnarray} \beta & = &(2\phi
\sin^2\theta_{12}\cos^2\theta_{13}+2q\phi\sin^2\theta_{13}-\phi)t
\nonumber \\& &  +\tan^{-1}\frac{\cos
2\theta_{12}\cos^2\theta_{13}\sin\phi t-\sin^2\theta_{13}\sin
(2q-1)\phi t} {\cos^2\theta_{13}\cos\phi t
+\sin^2\theta_{13}\cos(2q-1)\phi t}
\end{eqnarray}
If we take $q$ to be a rational number and  $t$ to be the cyclic
period, namely $\displaystyle t = \frac{2 \pi}{\omega_1 -
\omega_2}$, we recover the result in ref \cite{blasone}. However,
from the formula in ref \cite{blasone}, it is difficult to deduce
the mixing angles even if we can measure the Berry phase because
the formula involves too many unknowns. Clearly, our explicit
formula in eq(\ref{cp}) provides in principle a better means of
deducing the mixing angles and, more importantly, the CP-violating
phase through the measurement of the noncyclic geometric phases at
several different times and then solving the resulting
simultaneous equations. If necessary, errors in the measurement
can also be reduced by using some form of least square fit. Since
three flavor mixing is very important in CP violation, our formula
offers an invaluable tool for resolving the issue through the
measurement of geometric phase.

For completeness, we have also computed the other eight possible
components and their noncyclic geometric phase.  These results are
summarized as follows.
\begin{eqnarray}
\beta_{e \mu} & = & \omega_1 t \sin^2 \theta_{13} \sin^2
\theta_{23}(\cos( 2 \delta) - 1) - \phi t - 2 q \phi t \nonumber
\\ & & + 2 q \phi t \cos^2 \theta_{13} \sin^2 \theta_{23} + 2 \phi
t (\cos \theta_{12} \cos \theta_{23} - \cos \delta \sin
\theta_{12} \sin \theta_{13} \sin \theta_{23})^2 \nonumber \\ & &
\mbox{\hspace{5cm}} - \sin^2 \delta \sin^2 \theta_{12} \sin^2
\theta_{13} \sin^2 \theta_{23} \nonumber \\
 & &
+ {\footnotesize \tan^{-1} \frac{\left\{\begin{array}{l}\sin
\theta_{13} \sin \theta_{23} ( \sin(\phi t + \delta) - \sin^2
\theta_{12} \sin \zeta_- \\\mbox{\hspace{2cm}} - \cos^2
\theta_{12} \sin \zeta_+  )
\\ \mbox{\hspace{2cm}} - \cos \theta_{23} \cos(2 q \phi t) \sin (\phi t) \sin (2
\theta_{12}) \end{array} \right\}
 }{\left\{ \begin{array}{l} \sin \theta_{13} \sin \theta_{23} ( \cos(\phi t +
\delta) - \sin^2 \theta_{12} \sin \zeta_- \\\mbox{\hspace{2cm}} -
\cos^2 \theta_{12} \sin \zeta_+ ) \\\mbox{\hspace{2cm}} - \cos
\theta_{23} \sin(2 q \phi t) \sin (\phi t) \sin (2 \theta_{12})
\end{array} \right\}} } \\
& & \nonumber \\ & & \nonumber \\ & & \nonumber \\
\beta_{e \tau} & = & \omega_1 t \sin^2 \theta_{13} \sin^2
\theta_{23}(\cos( 2  \delta) - 1) - \phi t - 2 q \phi t \nonumber
\\ & & + 2 q \phi t \cos^2 \theta_{13} \sin^2 \theta_{23} + 2 \phi
t (\cos \theta_{12} \cos \theta_{23} - \cos \delta \sin
\theta_{12} \sin \theta_{13} \sin \theta_{23})^2 \nonumber
\\ & & \mbox{\hspace{5cm}} - \sin^2 \delta \sin^2 \theta_{12} \sin^2
\theta_{13} \sin^2 \theta_{23} \nonumber \\ & & +{\footnotesize
\tan^{-1} \frac{\left\{\begin{array}{l}\sin \theta_{13} \cos
\theta_{23} ( \sin(\phi t + \delta) - \sin^2 \theta_{12} \sin
\zeta_- \\\mbox{\hspace{2cm}} - \cos^2 \theta_{12} \sin \zeta_+  )
\\ \mbox{\hspace{2cm}}- \sin \theta_{23} \cos(2 q \phi t) \sin (\phi t) \sin (2
\theta_{12}) \end{array} \right\}
 }{\left\{ \begin{array}{l} \sin \theta_{13} \cos \theta_{23} ( \cos(\phi t +
\delta) - \sin^2 \theta_{12} \sin \zeta_- \\\mbox{\hspace{2cm}} -
\cos^2 \theta_{12} \sin \zeta_+ ) \\\mbox{\hspace{2cm}} - \sin
\theta_{23} \sin(2 q \phi t) \sin (\phi t) \sin (2 \theta_{12})
\end{array} \right\}}} \\
& & \nonumber \\ & & \nonumber \\ & & \nonumber \\
\beta_{\mu e} & = & \omega_1 t \sin^2 \theta_{13} (\cos(2 \delta)
- 1) - \phi t
 + 2 q \phi t \sin^2 \theta_{13} \cos(2 \delta)
 + 2 \sin^2 \theta_{12} \cos^2 \theta_{12} \phi t
 \nonumber \\ & &
+ {\footnotesize \tan^{-1} \frac{\left\{\begin{array}{l} 
\sin \theta_{13} \sin \theta_{23} 
\bigg( \sin^2 \theta_{12} \sin(\phi t + \delta)
\\ \mbox{\hspace{2cm}} - \cos^2 \theta_{12} \sin (\phi t - \delta)
 - \sin \zeta_- \bigg) \\ \mbox{\hspace{2cm}} - \cos \theta_{23} \sin (\phi
t ) \sin(2 \theta_{12})\end{array} \right\}
 }{\left\{  \begin{array}{l}
\sin \theta_{13} \sin \theta_{23} \bigg( \sin^2 \theta_{12}
\sin(\phi t + \delta) \\ \mbox{\hspace{2cm}} + \cos^2 \theta_{12}
\sin (\phi t - \delta)
\\ \mbox{\hspace{2cm}} + \cos \zeta_- \bigg) \end{array} \right\}}} \\
& & \nonumber \\ & & \nonumber \\ & & \nonumber \\
\beta_{\mu \mu} & = & \omega_1 t \sin^2 \theta_{13} \sin^2
\theta_{23}(\cos(2 \delta) - 1) + \phi t \nonumber
\\ & & + 2 q \phi t \cos^2 \theta_{13} \sin^2 \theta_{23} + 2 \phi
t (\cos \theta_{12} \cos \theta_{23} - \cos \delta \sin
\theta_{12} \sin \theta_{13} \sin \theta_{23})^2 \nonumber \\& &
\mbox{\hspace{5cm}} - \sin^2 \delta \sin^2 \theta_{12} \sin^2
\theta_{13} \sin^2 \theta_{23} \nonumber \\ & & + {\footnotesize
\tan^{-1} \frac{\left\{\begin{array}{l} \sin^2 \theta_{13} \sin^2
\theta_{23} \bigg( \cos^2 \theta_{12} \sin(\phi t - 2 \delta)\\
\mbox{\hspace{2cm}} - \sin^2 \theta_{12} \sin(\phi t + 2
\delta)\bigg)
\\ \mbox{\hspace{2cm}}  - \cos^2 \theta_{23} \cos(2 \theta_{12})
\sin(\phi t) \\ \mbox{\hspace{2cm}} - \cos^2 \theta_{13} \sin^2
\theta_{23} \sin(2 q -1)\phi t \\\mbox{\hspace{2cm}}  + \cos
\delta \sin \theta_{13} \sin(2 \theta_{12}) \sin(2 \theta_{23})
\sin (\phi t)
 \end{array} \right\}
 }{\left\{ \begin{array}{l}
 \sin^2 \theta_{13}
\sin^2 \theta_{23} \bigg( \cos^2 \theta_{12} \sin(\phi t - 2
\delta) \\ \mbox{\hspace{2cm}} + \sin^2 \theta_{12} \sin(\phi t +
2 \delta) \bigg) \\
 \mbox{\hspace{2cm}} - \cos^2 \theta_{23}  \cos(\phi t) \\
 \mbox{\hspace{2cm}} - \cos^2 \theta_{13} \sin^2 \theta_{23} \cos(2
q -1)\phi t \\ \mbox{\hspace{2cm}} + \sin \delta \sin \theta_{13}
\sin(2 \theta_{12}) \sin(2 \theta_{23}) \sin (\phi t)
\end{array} \right\}} }\\
& & \nonumber \\ & & \nonumber \\ & & \nonumber \\
& & \nonumber \\
\beta_{\mu \tau} & = & \omega_1 t \sin^2 \theta_{13} \sin^2
\theta_{23}(\cos(2 \delta) - 1) - \phi t  \nonumber
\\ & & + 2 q \phi t \cos^2 \theta_{13} \sin^2 \theta_{23} + 2 \phi
t (\cos \theta_{12} \cos \theta_{23} - \cos \delta \sin
\theta_{12} \sin \theta_{13} \sin \theta_{23})^2 \nonumber \\& &
\mbox{\hspace{5cm}} - \sin^2 \delta \sin^2 \theta_{12} \sin^2
\theta_{13} \sin^2 \theta_{23} \nonumber \\ & &
+{\footnotesize \tan^{-1} \frac{\left\{\begin{array}{l}
\frac{1}{2} \sin(2 \theta_{23}) \bigg( \cos^2 \theta_{13} \sin(2 q
-1 )\phi t + \cos(2 \theta_{12}) \sin (\phi t) \\
\mbox{\hspace{2cm}} + \sin^2 \theta_{13} \cos^2 \theta_{12}
\sin(\phi t - 2 \delta) \\ \mbox{\hspace{2cm}} - \sin^2
\theta_{13} \sin^2 \theta_{12} \sin(\phi t + 2 \delta) \bigg)
\\ \mbox{\hspace{2cm}}+ \cos \delta \sin \theta_{13} \sin(2
\theta_{12})\cos (2 \theta_{23}) \sin(\phi t)
\end{array} \right\}
 }{\left\{ \begin{array}{l}
\frac{1}{2} \sin(2 \theta_{23}) \bigg( \cos^2 \theta_{13} \cos(2 q
-1 )\phi t - \cos (\phi t)
\\ \mbox{\hspace{2cm}} + \cos^2 \theta_{13} \cos^2 \theta_{12}
\cos(\phi t - 2 \delta) \\ \mbox{\hspace{2cm}} - \cos^2
\theta_{13} \sin^2 \theta_{12} \cos(\phi t + 2 \delta) \bigg)
\\ \mbox{\hspace{2cm}}+ \sin \delta \sin \theta_{13} \sin(2
\theta_{12})\cos (2 \theta_{23}) \sin(\phi t)
\end{array} \right\}}}\\ & & \nonumber \\  & & \nonumber \\
 & & \nonumber \\
\beta_{\tau e} & = & \omega_1 t \sin^2 \theta_{13} (\cos(2 \delta)
- 1) - \phi t
 + 2 q \phi t \sin^2 \theta_{13}\cos(2 \delta)
 + 2 \sin^2 \theta_{12} \cos^2 \theta_{12} \phi t \nonumber \\ & &
+ {\footnotesize \tan^{-1} \frac{\left\{\begin{array}{l} 
\sin \theta_{23} \sin(2 \theta_{12}) \sin(\phi t)
\\ \mbox{\hspace{2cm}} + \sin \theta_{13} \cos \theta_{23} \bigg(
\cos^2 \theta_{12} \sin(\phi t - \delta) \\ \mbox{\hspace{2cm}} +
\sin^2 \theta_{12} \sin(\phi t + \delta) + \sin \zeta_- \bigg)
\end{array} \right\}
 }{\left\{  \begin{array}{l}
\sin \theta_{13} \cos \theta_{23} \bigg( \cos^2 \theta_{12}
\cos(\phi t - \delta) \\ \mbox{\hspace{2cm}} + \sin^2 \theta_{12}
\cos(\phi t + \delta) + \cos \zeta_- \bigg)
\end{array} \right\}}}\\ & & \nonumber \\  & & \nonumber \\
 & & \nonumber \\
\beta_{\tau \mu} & = & \omega_1 t \sin^2 \theta_{13} \sin^2
\theta_{23}(\cos(2 \delta) - 1) - \phi t  \nonumber
\\ & & + 2 q \phi t \cos^2 \theta_{13} \sin^2 \theta_{23} + 2 \phi
t (\cos \theta_{12} \cos \theta_{23} - \cos \delta \sin
\theta_{12} \sin \theta_{13} \sin \theta_{23})^2 \nonumber \\& &
\mbox{\hspace{5cm}} - \sin^2 \delta \sin^2 \theta_{12} \sin^2
\theta_{13} \sin^2 \theta_{23} \nonumber \\ & &
+{\footnotesize \tan^{-1} \frac{\left\{\begin{array}{l}
\frac{1}{2} \sin(2 \theta_{23}) \bigg( - \cos^2 \theta_{13} \sin(2
q -1 )\phi t + \cos(2 \theta_{12}) \sin (\phi t) \\
\mbox{\hspace{2cm}} + \sin^2 \theta_{13} \cos^2 \theta_{12}
\sin(\phi t - 2 \delta) \\ \mbox{\hspace{2cm}} + \sin^2
\theta_{13} \sin^2 \theta_{12} \sin(\phi t + 2 \delta) \bigg)
\\ \mbox{\hspace{2cm}}+ \cos \delta \sin \theta_{13} \sin(2
\theta_{12})\cos (2 \theta_{23}) \sin(\phi t)
\end{array} \right\}
 }{\left\{ \begin{array}{l}
\frac{1}{2} \sin(2 \theta_{23}) \bigg( \cos^2 \theta_{13} \cos(2 q
-1 )\phi t - \cos (\phi t)
\\ \mbox{\hspace{2cm}} + \cos^2 \theta_{12} \sin^2 \theta_{13}
\cos(\phi t - 2 \delta) \\ \mbox{\hspace{2cm}}+ \sin^2 \theta_{12}
\sin^2 \theta_{13} \cos(\phi t + 2 \delta) \bigg)
\\ \mbox{\hspace{2cm}}+ \sin \delta \sin \theta_{13} \sin(2
\theta_{12})\cos (2 \theta_{23}) \sin(\phi t)
\end{array} \right\}}}\\ & & \nonumber \\  & & \nonumber \\
 & & \nonumber \\
\beta_{\tau \tau} & = & \omega_1 t \sin^2 \theta_{13} \sin^2
\theta_{23}(\cos(2 \delta) - 1) - \phi t \nonumber
\\ & & + 2 q \phi t \cos^2 \theta_{13} \sin^2 \theta_{23} + 2 \phi
t (\cos \theta_{12} \cos \theta_{23} - \cos \delta \sin
\theta_{12} \sin \theta_{13} \sin \theta_{23})^2 \nonumber \\& &
\mbox{\hspace{5cm}} - \sin^2 \delta \sin^2 \theta_{12} \sin^2
\theta_{13} \sin^2 \theta_{23} \nonumber \\ & & - {\footnotesize
\tan^{-1} \frac{\left\{\begin{array}{l} \sin^2 \theta_{13} \sin^2
\theta_{23} \bigg( \cos^2 \theta_{12} \sin(\phi t - 2 \delta)\\
\mbox{\hspace{2cm}} - \sin^2 \theta_{12} \sin(\phi t + 2
\delta)\bigg)
\\ \mbox{\hspace{2cm}}  + \cos^2 \theta_{23} \cos(2 \theta_{12})
\sin(\phi t) \\ \mbox{\hspace{2cm}} + \cos^2 \theta_{13} \cos^2
\theta_{23} \sin(2 q -1)\phi t \\\mbox{\hspace{2cm}}  + \cos
\delta \sin \theta_{13} \sin(2 \theta_{12}) \sin(2 \theta_{23})
\sin (\phi t)
 \end{array} \right\}
 }{\left\{ \begin{array}{l}
 \sin^2 \theta_{13}
\sin^2 \theta_{23} \bigg( \cos^2 \theta_{12} \sin(\phi t - 2
\delta) \\ \mbox{\hspace{2cm}} + \sin^2 \theta_{12} \sin(\phi t +
2 \delta) \bigg) \\
 \mbox{\hspace{2cm}} + \sin^2 \theta_{23}  \cos(\phi t) \\
 \mbox{\hspace{2cm}} + \cos^2 \theta_{13} \cos^2 \theta_{23} \cos(2
q -1)\phi t \\ \mbox{\hspace{2cm}} - \sin \delta \sin \theta_{13}
\sin(2 \theta_{12}) \sin(2 \theta_{23}) \sin (\phi t)
\end{array} \right\}} }
\end{eqnarray}
where $\zeta_\pm = (2 q \pm 1) \phi t - \delta$.

\section{Discussion and conclusion}\label{discuss}

Although we have restricted our computation to neutrino
oscillations, the results can be extended easily to the case of
neutron-anti-neutron oscillation \cite{mohapatra}. Under certain
circumstances, the measurement of geometric phase can be obtained
more robustly in experiments and this idea of extending the
noncyclic geometric phases of neutrino oscillation to $n-\bar{n}$
oscillation may provide an alternative experimental basis for
detecting baryon number violation. Moreover, our results holds for
oscillations of  any mixed state bosons, for example Kaons,
$\eta^\prime$ and so forth.

It is noteworthy to remark that Berry phase has recently been
shown to exhibit essentially fault-tolerant behavior in quantum
computation through NMR experiments\cite{jones}. In general, 
this fault-tolerant 
behavior holds for any geometric phase, be it adiabatic or 
non-adiabatic, cyclic or non-cylic.
In a similar context, it has
also been shown to be suitable for analyzing entangled quantum
states\cite{agarwal}.

In summary, we have calculated the non-cyclic geometric phases
with both two-flavor and three-flavor mixing for the neutrino
oscillations. If we set the time of measurement to the period of
the oscillation, we recover the previous results found in ref
\cite{blasone}. Thus, our formulae naturally generalize the
results for the Berry phase\cite{blasone}. Finally, our formulae
could have a potential application for determining the mixing
angles of oscillating neutrinos and the CP violating phase.

\end{document}